\documentclass[twocolumn,
superscriptaddress,
amsmath,amssymb,
aps,
prl,
]{revtex4-1}

\usepackage{graphicx, color}
\usepackage{dcolumn}
\usepackage{bm}
\usepackage{ulem}

\usepackage[mathlines]{lineno}

\begin{document}

\title{Dynamic Magnetic Pair-Density Function of a One-Dimensional Ferromagnet}

\author{Shin-ichi Shamoto}
\email[Corresponding author:\ ]{shamoto-shinichi@rada.or.jp}
\affiliation{Neutron Application Promotion Group, Quantum Radiation Application Development Association (q-RADA), \\Tokai 319-1106, Japan}
\affiliation{School of Science and Technology, Meiji University, Kawasaki, Kanagawa 214-8571 Japan}
\affiliation{Department of Physics, National Cheng Kung University, Tainan 70101, Taiwan}

\date{\today}

\begin{abstract}
The dynamic magnetic pair-density function (DymPDF) $D_{\rm M}(r, E)$ is derived by extending the static magnetic pair distribution function (mPDF) to finite energy transfer. The analytical DymPDF of a one-dimensional Heisenberg ferromagnet is obtained from the magnon dispersion and compared with simulations performed using {\textsc SpinW}. Excellent agreement is achieved for energy dependence of the nearest-neighbor  spin-pair correlation, demonstrating that the DymPDF changes sign at the magnon-mode transition occurring at one-half of the maximum magnon energy. The $r$-dependence of DymPDF at low energy is also reproduced with a model including magnon correlation length and Fourier-termination effects. These results establish the theoretical foundation of DymPDF analysis for investigating local spin dynamics in magnetic materials.

Subject Areas: dynamic magnetic pair-density function, one-dimensional ferromagnet, inelastic neutron scattering, real-space spin dynamics, magnon mode transition
\end{abstract}
\maketitle

The Fourier-transform analysis has been widely used to study local structures of liquid and amorphous materials, even disorders in crystalline materials as pair distribution function (PDF) analysis by total neutron scattering and x-ray scattering \cite{Egami2019,EgamiBillinge}. The magnetic version has been developed to study the local magnetic structure of frustrated magnets as magnetic pair distribution function (mPDF) analysis \cite{Frandsen2014, Kodama2021}. 

Owing to recent advances in high-intensity pulsed neutron and synchrotron x-ray sources, Fourier-transform analyses can now be performed for both static and dynamic correlations. 
Studying the local dynamics is essential for understanding the physical properties of clustered materials at the nanoscale level. For instance, local lattice dynamics have been examined using the Fourier-transform of the lattice dynamic structure factor, $S_{\rm L}(Q, E)$, referred as dynamic pair density function (DyPDF) analysis \cite{Egami2019}. This analysis has revealed local phonon properties in various materials. They have been observed in silica glass\cite{Hannon1992}, high-$T_{c}$ cuprate \cite{Arai1995}, relaxor ferroelectrics\cite{Dmowski2008}, and the superfluid of $^{4}$He\cite{Dmowski2017}. A technical review of DyPDF is also available to check the worldwide availability of inelastic neutron scattering (INS) spectrometers \cite{Acosta2023}. 
The magnetic version is the dynamic magnetic pair-density function (DymPDF) analysis \cite{Iida2022}, which is developed using the state-of-the-art inelastic neutron scattering spectrometer, 4SEASONS \cite{Kajimoto2011, Nakamura2009}, at the Materials and Life Science Experimental Facility (MLF) at the Japan Proton Accelerator Research Complex (J-PARC). 
However, only an experimental part of the formalism of the DymPDF $D_{\rm M}(r, E)$ has been shown in ref. \citenum{Iida2022}. To improve understanding of the formalism, we derive the expression of the DymPDF by extending the mPDF formalism to finite energy and apply it to a one-dimensional (1D) ferromagnet as a representative example.

First, we define the coherent magnetic structure factor $S_{\rm M}(Q)$ using the coherent magnetic differential cross-section as
\begin{equation}
\left(\frac{d \sigma}{d\Omega}\right)_{\rm M} 
= \frac{2}{3}NS(S+1)\left(\frac{\gamma r_{0}g}{2}\right)^{2} f_{\rm M}^{2}(Q) S_{\rm M}(Q),
\label{eq:1}
\end{equation}
where 2/3 factor comes from the powder-averaged angle-dependent term of magnetic elastic scattering, $[1-(\hat{\bf Q} \cdot \hat{\mathbf s})^2]_{av}$ \cite{Squires}; $N$ is number of spins in the system; $S$ is a spin quantum number; $\gamma = -1.913$ is the neutron gyromagnetic ratio in nuclear magnetons; the classical electron radius $r_{0} = 2.818 \times 10^{-15}$ m; $g$ is the Land\'e $g$-factor; $f_{\rm M}(Q)$ is a magnetic form factor.

Then, we review the static magnetic pair-distribution function (mPDF) \cite{Frandsen2014, Kodama2021}. The static mPDF is written in polar coordinates as follows.
\begin{eqnarray}
S_{\rm M}(Q)-1 = \nonumber\\\frac{1}{N}\sum_{i\neq j}\Bigl[ A_{ij}\frac{\sin({Q r}_{ij})}{{Q r}_{ij}}
+B_{ij} \Big( \frac{\sin({Q r}_{ij})}{({Q r}_{ij})^{3}}-\frac{\cos({Q r}_{ij})}{({Q r}_{ij})^{2}}\Big) \Bigl],
\label{eq:2}
\end{eqnarray}    
where $A_{ij} = \langle S_{i}^{y} S_{j}^{y}\rangle$ and $B_{ij} = 2\langle S_{i}^{x} S_{j}^{x}\rangle - \langle S_{i}^{y} S_{j}^{y}\rangle$, which are integrated in energy; the subscripts $i$ and $j$ refer to individual magnetic moments $S_i$ and $S_j$ separated by a distance $r_{ij}$; $Q$ is the magnitude of the scattering vector, which is equivalent to $\kappa$ in ref. \citenum{Frandsen2014}.
The coordinate system for $A_{ij}$ and $B_{ij}$ in ref. \citenum{Frandsen2014} is locally defined for each spin pair through
\begin{eqnarray}
\hat {\bf x} = \frac{{\bf{r}}_{j}-{\bf{r}}_{i}}{|{\bf{r}}_{j}-{\bf{r}}_{i}|} \;{\rm and}\; {\bf \hat y} = \frac{{\mathbf{S}}_{i}- {\bf \hat x}({\mathbf {S}}_{i}\cdot \bf{\hat x})}{|{\mathbf{S}}_{i}-{\bf \hat x}({\mathbf {S}}_{i}\cdot \bf{\hat x})|},
\label{eq:3}
\end{eqnarray}  
where unit vector $\hat {\bf x}$ is along the bond direction between $i$ and $j$ sites; ${\mathbf{S}}_{i}$ is a classical spin vector at site $i$.

The corresponding real-space mPDF is written in polar coordinates as follows \cite{Frandsen2014}.
\begin{eqnarray}
f(r) = \frac{2}{\pi} \int_{0}^{\infty} dQ [S_{\rm M}(Q)-1]Q\sin(Q r)\nonumber\\
=\frac{1}{N}\frac{3}{2S(S+1)}\sum_{i\neq j}\Big\{\frac{A_{ij}}{r}\delta (r-r_{ij})\nonumber\\
+B_{ij}\frac{r}{r_{ij}^3}[1-\Theta(r-r_{ij})]\Big\},
\label{eq:4}
\end{eqnarray}  
where the Heaviside step function $\Theta(r-r_{ij})$ = 0 for $r < r_{ij}$ or 1 for  $r > r_{ij}$. In the mPDF analysis, the spin components of $S_{i}^{x,y}$ and $S_{j}^{x,y}$ that are normal to the bond vector ${\bf r}_{ij}$ contribute to the peak intensity term $A_{ij}$ \cite{Frandsen2014}. The magnetic Bragg peak intensity is zero when the unit spin vector $\hat{\mathbf s}$ and the unit scattering vector $\hat{\bf Q}$ are parallel. These specific scattering conditions can be averaged for uniformly distributed powdered magnets, resulting in the factor of 3/2. The same powder averaging can be applied to the inelastic neutron scattering case.

We now extend the mPDF formalism to the DymPDF at finite energy $E$.
We define the coherent dynamic magnetic structure factor $S_{\rm M}(Q, E)$ as
\begin{equation}
\left(\frac{d^2 \sigma}{d\Omega dE}\right)_{\rm M} 
= \frac{4}{3}\frac{k_{f}}{k_{i}}N\left(\frac{\gamma r_{0}g}{2}\right)^{2} f_{\rm M}^{2}(Q) S_{\rm M}(Q, E),
\label{eq:5}
\end{equation}
where  4/3 factor comes from the powder-averaged angle-dependent term of magnetic inelastic neutron scattering, $[1+(\hat{\bf Q} \cdot \hat{\mathbf s})^2]_{av}$ \cite{Squires, Shamoto2018, Shamoto2023}; $k_{i}$ and $k_{f}$ are incident and final wave numbers, respectively.
Then,  the dynamic magnetic structure factor can be written in polar coordinates as
\begin{eqnarray}
\frac{S_{\rm M}(Q, E)}{\langle n(E)+1\rangle}-1 =
 \frac{1}{N}\sum_{i\neq j}\Bigl[ A_{ij}(E)\frac{\sin({Q r}_{ij})}{{Q r}_{ij}}\nonumber\\
+B_{ij}(E) \Big( \frac{\sin({Q r}_{ij})}{({Q r}_{ij})^{3}}-\frac{\cos({Q r}_{ij})}{({Q r}_{ij})^{2}}\Big) \Bigl],
\label{eq:6}
\end{eqnarray}    
where $\langle n(E)+1\rangle$ is the Bose factor; $A_{ij}(E)$ and $B_{ij}(E)$ depend on the magnon dispersion at the energy $E$, which will be calculated in a 1D ferromagnet case. The extension of Eq. \ref{eq:2} to Eq. \ref{eq:6} generally corresponds to the introduction of time correlation as $A_{ij}(t) = \langle S_{i}^{y}(0) S_{j}^{y}(t)\rangle$; $B_{ij}(t) = 2\langle S_{i}^{x}(0) S_{j}^{x}(t)\rangle - \langle S_{i}^{y}(0) S_{j}^{y}(t)\rangle$, which are Fourier transformed to $A_{ij}(E)$ and $B_{ij}(E)$. This general extension is an essential part of this work, leading to the wide applicability of current formalism.

The corresponding real-space DymPDF is written in the polar coordinates as
\begin{eqnarray}
D_{\rm M}(r, E) = f(r, E) \nonumber\\
= \frac{2}{\pi} \int_{0}^{\infty} dQ \left[\frac{S_{\rm M}(Q, E)}{\langle n(E)+1\rangle}-1\right]Q \sin(Q r) \nonumber\\
=\frac{1}{N}\frac{3}{4}\sum_{i\neq j}\Big\{\frac{A_{ij}(E)}{r}\delta (r-r_{ij})\nonumber\\
+B_{ij}(E)\frac {r}{r_{ij}^3}[1-\Theta(r-r_{ij})]\Big\}.
\label{eq:7}
\end{eqnarray}  
This DymPDF equation can be generally applied for standard magnets.

The powder-averaged DymPDF is experimentally obtained as follows \cite{Iida2022}.
 The dynamic magnetic structure factor $S_{\rm M}(Q, E)$ in polar coordinates is Fourier-transformed to the DymPDF $D_{\rm M}$($r, E$) based on the following equations, after being divided by the Bose factor $\langle n(E)+1\rangle$.
\begin{eqnarray}
\frac{S_{\rm M}(Q, E)}{\langle n(E)+1\rangle}-1 = \nonumber\\
\frac{S(Q, E)-A(E)(Q-Q_{\rm max})^{2}-S_{0}(Q_{\rm max},E)}{(4/3)N(\gamma r_{0}g/2)^{2} f_{\rm M}^{2} (Q)\langle n(E)+1\rangle}\;,
\label{eq:8}
\end{eqnarray}
where we set $Q_{\rm max}$ to be 5 \AA$^{-1}\;$; a parameter $A(E)$ is calculated at $E$ by least squares fitting to minimize the integral of $[S(Q, E)-A(E)(Q-Q_{\rm max})^{2}-S_{0}(Q_{\rm max}, E)]$ from $Q_{\rm min}$ to $Q_{\rm max}$. This subtraction usually removes the phonon components while correcting the oscillation balance of $S_{\rm M}(Q, E)/\langle n(E)+1\rangle$ around unity. $S_{0}(Q_{\rm max}, E)$ is determined at $Q = Q_{\rm max}$ so that the numerator is zero on the right-hand side of the equation. The same program in {\textsc Utsusemi}\cite{Utsusemi} is used to obtain the simulation. In the program, $A(E)$ and $S_{0}(Q_{\rm max}, E)$ are treated as fitting parameters.
\begin{eqnarray}
D_{\rm M}(r, E) = \frac{2}{\pi} \int_{Q_{\rm min}}^{Q_{\rm max}}dQ\nonumber\\
\left[\frac{S_{\rm M}(Q, E)}{\langle n(E)+1\rangle}-1\right] w(Q) Q\sin(Qr),
\label{eq:9}
\end{eqnarray}
where $w(Q)$ is a window function $(Q_{\rm max}/\pi Q)\sin(\pi Q/Q_{\rm max})$ to suppress the diverging error near $Q_{\rm max}$.

\begin{figure}[ht] 
\includegraphics[width=7.5cm,clip]{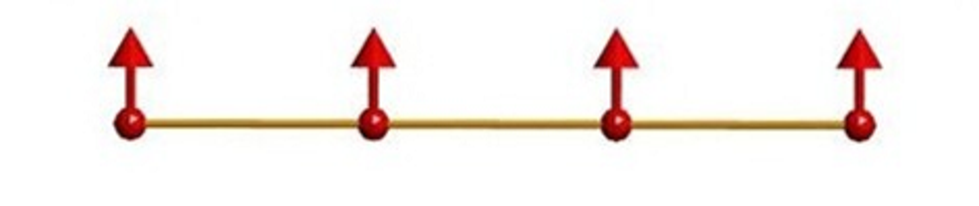}
\caption{A part of magnetic structures of one-dimensional (1D) ferromagnet with nearest-neighbor bond distance of $a$. 
}
\label{model}
\end{figure}
The dynamic magnetic scattering function of a 1D Heisenberg ferromagnet was calculated using the SpinW software package \cite{SpinW}, which is based on the linear spin-wave theory with Holstein-Primakoff approximation.  The spin Hamiltonian and the parameters are as follows.
\begin{eqnarray}
H = \sum_{i,j}J_{ij} \hat{\mathbf S}_{i} \cdot \hat{\mathbf S}_{j},
\label{eq:10}
\end{eqnarray}
where $H$ is the spin Hamiltonian; $J_{ij}$ is the exchange parameter between $\hat{\mathbf S}_{i}$ and $\hat{\mathbf S}_{j}$; $\hat{\mathbf S}_{i}$ and $\hat{\mathbf S}_{j}$ are spin operators at $i$-$th$ and $j$-$th$ spins, respectively. We consider only the simplest case, which involves nearest-neighbor interaction with $ J_{ij}$ = $ J_{a}$ = $-$1 meV and $S$ = 1.  The nearest-neighbor bond distance $a$ is 3 \AA. The spin model structure of 1D ferromagnet is shown in Fig. \ref{model}.  
 
\begin{figure}[ht] 
\includegraphics[width=8.0cm,clip]{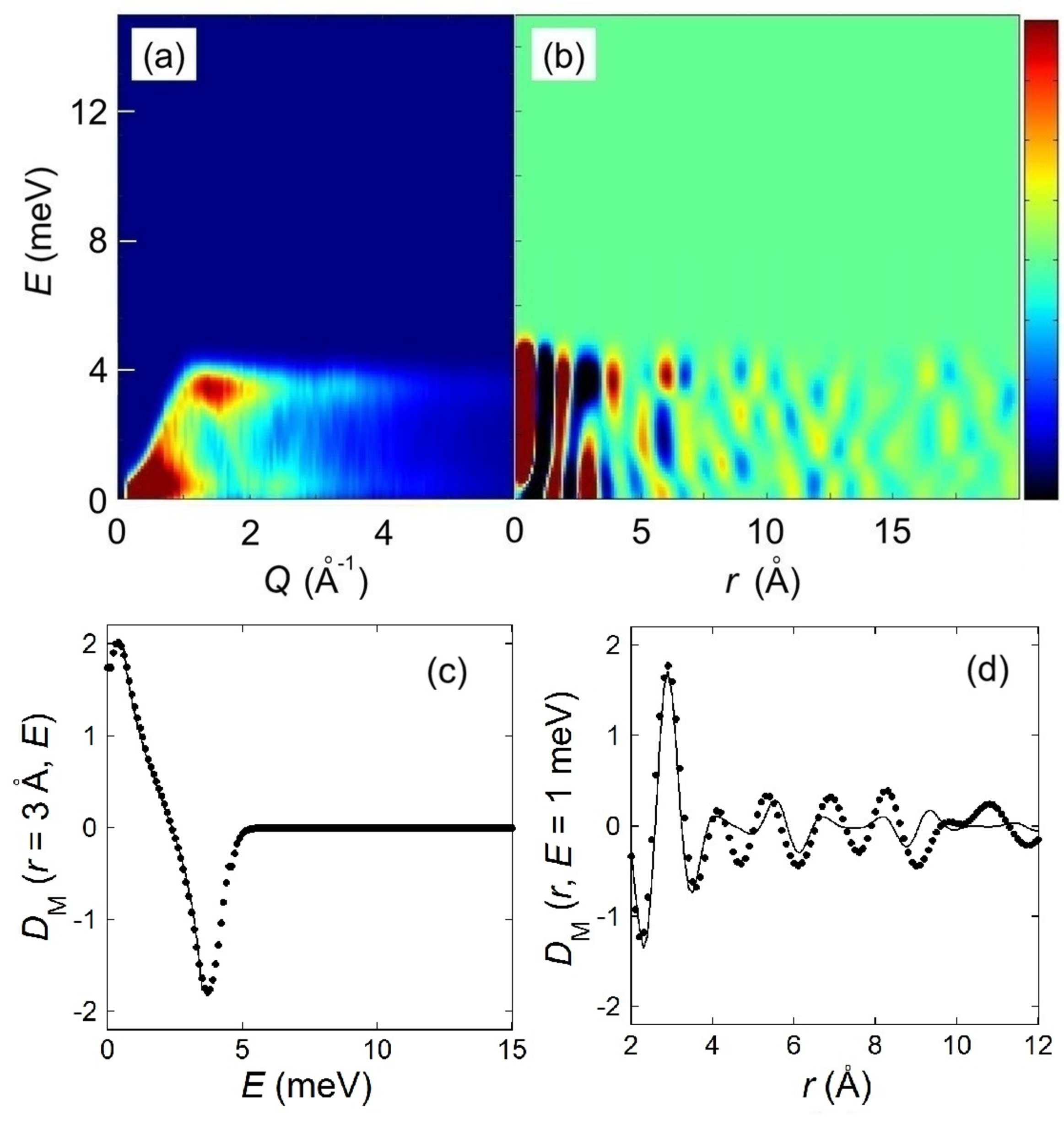}
\caption{(a) Powder-averaged magnetic dynamic structure factor of the one-dimensional ferromagnet calculated using SpinW. (b) Corresponding DymPDF map. (c) Energy dependence of the nearest-neighbor DymPDF. (d) Real-space DymPDF at 1 meV. Solid curves are fits using Eqs. (22) and (23).
}
\label{1D}
\end{figure}

The powder-averaged dynamic structure factor of the 1D ferromagnet calculated using {\textsc SpinW} is shown in Fig. \ref{1D}(a). The DymPDF patterns were calculated using ‘Cmm.DymPDFFromSpinWText‘ command in ‘Utsusemi‘ \cite{Utsusemi}. For the 1D ferromagnet, the maximum magnon energy is 4 meV. 
The ferromagnetic magnon dispersion, with the chain direction along the $x$-axis is expressed as

\begin{eqnarray}
E({\bf k}) = 2|J_a|S[1-\cos (k_{x}a)], 
\label{eq:11}
\end{eqnarray}    
where $J_{a}$ is the ferromagnetic nearest-neighbor exchange parameter; $a$ is the bond distance between the nearest-neighbor magnetic moments; ${\bf k}=(k_{x}, k_{y}, k_{z})$ is the wave number vector in Cartesian coordinates, where $k_y$ and $k_z$ are the components normal to the chain. The wave number vector ${\bf k}$ can be written in polar coordinates as $k_{x}=Q\cos{\theta}, k_{y}=Q\sin{\theta}\cos{\phi}, k_{z}=Q \sin{\theta} \sin{\phi}$, which lead to $Q^{2}=k_{x}^{2}+k_{y}^{2}+k_{z}^{2}$.
 
Before calculating $A_{ij}(E)$ and $B_{ij}(E)$, we introduce the following energy-resolved spin-correlation tensor $C_{ij}^{\alpha \beta}(E)$,

\begin{eqnarray}
C_{ij}^{\alpha \beta}(E) = \sum_{n} \langle 0|S_{i}^{\alpha}|n \rangle  \langle n|S_{j}^{\beta}|0 \rangle \delta (E-E_{n}),
\label{eq:12}
\end{eqnarray}
where $\alpha, \beta$ are components of $x, y, z$ in Cartesian coordinates at the sites $i$ and $j$.   

It satisfies the sum rule
\begin{eqnarray}
\int_{-\infty}^{\infty}dE C_{ij}^{\alpha \beta}(E) = \langle 0|S_{i}^{\alpha}S_{j}^{\beta}|0 \rangle.
\label{eq:13}
\end{eqnarray}
The original mPDF static coefficients $A_{ij}$ and $B_{ij}$ can be written as 
\begin{eqnarray}
A_{ij} = \int_{-\infty}^{\infty}dE C_{ij}^{yy}(E) = \langle S_{i}^{y}S_{j}^{y} \rangle,\nonumber\\
B_{ij} = \int_{-\infty}^{\infty}dE [2C_{ij}^{xx}(E) - C_{ij}^{yy}(E)] \nonumber\\
=  2\langle S_{i}^{x}S_{j}^{x} \rangle - \langle S_{i}^{y}S_{j}^{y} \rangle.
\label{eq:14}
\end{eqnarray}

Then energy-resolved $A_{ij}(E)$ can be written using the $C_{ij}^{\alpha \beta}(E)$ as
\begin{eqnarray}
A_{ij}(E) = C_{ij}^{yy}(E),
\label{eq:15}
\end{eqnarray}  
where  $C_{ij}^{xx}(E) = C_{ij}^{yy}(E)$ and $C_{ij}^{zz}\sim 0$ for an isotropic ferromagnet with the ordered moment along the $z$ axis. Here, the $x$ axis is defined locally along the bond direction ($\hat r_{ij} = \hat x$), while $y$ is perpendicular to both the bond and the ordered moment when the ordered moment direction is along $z$. 

The $B_{ij}(E)$ in isotropic one-magnon 1D ferromagnet for the chain along $x$ becomes
\begin{eqnarray}
B_{ij}(E) = 2 C_{ij}^{xx}(E)-C_{ij}^{yy}(E)= C_{ij}^{yy}(E) = A_{ij}(E).
\label{eq:16}
\end{eqnarray}  
Based on the 1D ferromagnet dispersion in Eq. \ref{eq:11}, the energy depends only on the wave-vector component along the chain, $k_x$. We therefore distinguish the intrinsic one-dimensional magnon dispersion from the three-dimensional momentum vector measured in the neutron scattering experiment,

For a spin pair separated along the chain, $r_{ij}=|x_j-x_i|$, the energy-resolved correlation function is obtained from the $k_x$ sum,
\begin{eqnarray}
A_{ij}(E)&=&\frac{S}{2N}\sum_{k_x}e^{ik_xr_{ij}}\delta(E-E_{k_x})\nonumber\\
&=&\frac{S}{2N}\frac{Na}{2\pi}\sum_{\nu=\pm}\frac{e^{ik_\nu r_{ij}}}{\left|\partial E/\partial k_x\right|_{k_\nu}}\nonumber\\
&=&\frac{\cos[q(E)r_{ij}]}{4\pi |J_a|\sqrt{\dfrac{E}{|J_a|S}\left(1-\dfrac{E}{4|J_a|S}\right)}},
\label{eq:17}
\end{eqnarray}
where $\delta(E-E_{k_x})=\sum_{\nu}\delta(k-k_{\nu})/\left|\partial E/\partial k_x\right|_{k_\nu}$; $k_{\nu}=\pm q(E)$; the magnon group velocity $\partial E/\partial k_x$ = $2|J_{a}|Sa \sin{(k_{x}a)}$, whereas $\partial E/\partial k_{y}$ = $\partial E/\partial k_{z}$ = 0; $k_{\nu}$ are the solutions of $E=E_{k_{x}}$; $1/|\partial E/\partial k_{x}|$ represents the 1D magnon density of states. The magnon group velocity vanishes at the band edges, producing the one-dimensional van Hove divergences at $E=0$ and $E=4|J_a|S$.
At fixed energy, the two chain-direction solutions are
\begin{eqnarray}
k_x=\pm q(E),\qquad q(E)=\frac{1}{a}\arccos\Big(1-\frac{E}{2|J_a|S}\Big),
\label{eq:18}
\end{eqnarray}
for $0<E<4|J_a|S$, while the transverse components satisfy $k_y^2+k_z^2=Q^2-q(E)^2$. Hence the powder-averaged scattering has support only for $Q\ge q(E)$. Although the magnon dispersion is one-dimensional and depends only on the chain wave vector $k_{x}$, the neutron scattering experiment measures the three-dimensional momentum transfer $\bf k$. The transverse components $k_{y}$ and $k_{z}$ therefore contribute to the powder-averaged phase space without modifying the magnon energy. Consequently, the constant-energy manifold forms a cylindrical surface around the chain axis, leading to the condition $Q \geq q(E)$.
 
Substitution of Eq. \ref{eq:17} into Eq. \ref{eq:7} gives
\begin{eqnarray}
D_{\rm M}(r,E)&\approx&\frac{3}{4N}\sum_{i\ne j}\frac{\cos[q(E)r_{ij}]}{4\pi |J_a|\sqrt{\dfrac{E}{|J_a|S}\left(1-\dfrac{E}{4|J_a|S}\right)}}\nonumber\\
&&\times\left\{\frac{\delta(r-r_{ij})}{r}+\frac{r}{r_{ij}^3}[1-\Theta(r-r_{ij})]\right\},
\label{eq:19}
\end{eqnarray}
where $B_{ij}(E)=A_{ij}(E)$ is used for the isotropic one-magnon ferromagnet.

For the nearest-neighbor, $r_{ij}=a$. Therefore,
\begin{eqnarray}
\cos[q(E)a]=1-\frac{E}{2|J_a|S}.
\label{eq:20}
\end{eqnarray}
Then the former term of Eq.\ref{eq:19} at $r = a$ is
\begin{eqnarray}
D_{\rm M}^{(A)}(a,E)&\approx&\frac{3}{4}\frac{1-\dfrac{E}{2|J_a|S}}{4\pi |J_a|a\sqrt{\dfrac{E}{|J_a|S}\left(1-\dfrac{E}{4 |J_a|S}\right)}}.
\label{eq:21}
\end{eqnarray}
The Heaviside function in Eq. \ref{eq:19} is approximated by Fermi Dirac function with a broadening parameter $\Gamma$ based on ref. \citenum{Frandsen2014}, which smooths the Heaviside function as a finite resolution effect.

Introducing $E_{\rm max}=4|J_a|S$ and $E_{\rm t}=2|J_a|S=E_{\rm max}/2$, the fitting function becomes
\begin{eqnarray}
D_{\rm M}(a,E)\approx A_{0}\frac{1-E/(E_{\rm max}/2)}{\sqrt{(E/E_{\rm max})(1-E/E_{\rm max})}} \nonumber\\
+B_{0}\Big[1+\tanh \Big(\frac{E-E_{\rm max}/2}{\Gamma}\Big)\Big].
\label{eq:22}
\end{eqnarray}
The sign reversal occurs at $E_{\rm max}/2= E_{\rm t}= 2|J_a|S$. The fitting by Eq. \ref{eq:22} to Fig. \ref{1D}(c) gives $E_{\rm max}=3.93 \pm 0.04$ meV and $\Gamma=2.3 \pm 1.6$ meV, in excellent agreement with the simulation parameter $E_{\rm max}=4$ meV.

For the $r$-dependence, $r_{ij}=|n|a$, where $n$ is an integer. At $E = |J_a|S$ = 1 meV, $q(E)a=\pi/3$ in Eq. \ref{eq:19}, so that $2\cos[n\pi/3]=\{1,-1,-2,-1, ...\}$. To describe the finite $r$-resolution, the delta function is replaced by a Gaussian of width $\sigma$, and a correlation length $\xi$ is introduced. The finite $Q_{\rm max}$ produces the Fourier-termination function $K(r)=\sin(Q_{\rm max}r)/(Q_{\rm max}r)$. Neglecting the $B_{ij}(E)$ term and retaining four neighbor shells gives
\begin{eqnarray}
D_{\rm M}(r,|J_a|S)\approx \frac{1}{4\pi}\sqrt{\frac3 4}\frac{K(r)e^{-r/\xi}}{|J_a|}\frac{a}{r}\nonumber\\
\times \sum_{|n| \neq 0}2\cos\Big(\frac{n\pi}{3})\frac{1}{\sqrt{2\pi}\sigma}\exp\Big[-\frac{(r-na)^{2}}{2\sigma^{2}}\Big]\nonumber\\
\approx  \frac{1}{4\pi}\sqrt{\frac3 4}\frac{K(r)e^{-r/\xi}}{|J_a|\sqrt{2\pi}\sigma}\frac{a}{r} (e_1-e_2-2e_3-e_4),
\label{eq:23}
\end{eqnarray}
where $e_{n}=\exp[-(r-na)^2/(2\sigma^2)]$.
The factor $a/r$ retains the real-space $1/r$ dependence. The fitting gives $Q_{\rm max}$=4.85 $\pm$ 0.02 \AA$^{-1}$, $\sigma$=0.81 $\pm$ 0.06 \AA, and $\xi$=16 $\pm$ 13 \AA, consistent with the simulation parameters, $Q_{\rm max}$= 5 \AA$^{-1}$, $\sigma \sim \sqrt{2\ln(2)}\pi/Q_{\rm max} \sim $ 0.73 \AA.

In summary, we have established the theoretical formalism of the dynamic magnetic pair-density function (DymPDF) by extending the magnetic pair distribution function to finite energy transfer. The analytical solution for a one-dimensional Heisenberg ferromagnet demonstrates that the DymPDF is governed by both the one-dimensional magnon density of states and the real-space mode factor, leading to a sign reversal at the magnon-mode transition energy $E_{t} = 2|J_{a}|S$. Quantitative agreement with SpinW simulations confirms the validity of the formalism. The DymPDF provides a real-space approach for investigating local spin dynamics in magnetic materials and is expected to be applicable to nanomagnets, clustered magnets, frustrated magnets, and amorphous magnetic systems.
\\

\begin{acknowledgments}
We thank K. Kodama, H. Shamoto, L.-J. Chang, Y. Inamura, J.-H. Chung, Y. Yasui, K. Iida, and M. Nakamura for their valuable discussions and help. This work was supported by Grants-in-Aid for Scientific Research (C) (No. JP22K04678, JP25K08263) from the Japan Society for the Promotion of Science.  
\end{acknowledgments}

\end{document}